\documentclass[aps,pr,twocolumn]{revtex4}
\makeatletter

\newcommand{\Rmnum}[1]{\expandafter\@slowromancap\romannumeral #1@}
\makeatother
\usepackage{graphicx}
\usepackage{dcolumn}
\usepackage{bm}
\usepackage{CJK}
\usepackage{color}
\usepackage{amsfonts}
\usepackage{psfrag}
\usepackage{wrapfig}
\usepackage{subfigure}
\usepackage{makeidx}
\usepackage{multirow}
\usepackage{epsf}
\usepackage[colorlinks,linkcolor=red,anchorcolor=red,citecolor=blue,urlcolor=blue]{hyperref}
\usepackage{amsmath}
\usepackage{cases}
\usepackage{bm}

\begin{document}
\begin{CJK}{GBK}{song}
\title{Exact analytic spectra of rogue waves for Manakov equations}
\author{Shao-Chun Chen$^{1}$}
\author{Chong Liu$^{1,3,4}$}\email{chongliu@nwu.edu.cn}
\author{Nail Akhmediev$^{2}$}
\affiliation{$^1$School of Physics, Northwest University, Xi'an 710127, China}
\affiliation{$^2$Research School of Physics and Engineering, The Australian National University, Canberra, ACT 2600, Australia}
\affiliation{$^3$Shaanxi Key Laboratory for Theoretical Physics Frontiers, Xi'an 710127, China}
\affiliation{$^4$Peng Huanwu Center for Fundamental Theory, Xi'an 710127, China}
\begin{abstract}
The spectra of rogue waves of Manakov equations that exist in both focusing and defocusing regimes are derived in analytic form. These spectra are asymmetric during their whole expansion-contraction cycle. They have triangular shape at each side of the spectrum in the log scale. Such spectra are characterised by two quantities: the slopes of the spectra and the spectral jump at zero frequency.
We confirm our analytical results using numerical simulations.
\end{abstract}

\maketitle
\section{Introduction}

The notion of rogue waves (RW) have played a fundamental role in modeling extreme wave events \cite{RW09,pattern2009,Shrira10,theoryreview2017}.
This concept occurred to be useful in various branches of science including oceanography, hydrodynamics, plasma physics, and optics \cite{review2013,review2014,review2019}.
Mathematically, RWs are solutions of evolution equations that are localised both in time and in space. Examples are the rational solution of the nonlinear Schr\"odinger equation known as Peregrine solution \cite{Peregrine} and its higher-order analogs \cite {RW09}.

The degree of localisation of RWs in space evolves in time reaching its maximum at certain time. Correspondingly, the spectra of such solutions experience expansion-contraction cycle. In the particular case of the Peregrine solution, the spectrum is triangular and symmetric with equal slopes of the spectrum at each side \cite{Spectra2011-1,Spectra2011-2,RWO-1,RWO-2}. The spectra are the widest at the point of maximum localisation of the RW. Such spectra have been observed in experiments \cite{RWO-1,RWO-2}.

Rogue wave solutions have been found for many integrable models \cite{VRW-chap-2016,Rewiew-2017,PLA2012,ds-oyI,FB3,JS2013,Feng2016}. However, research on spectral properties of RWs remains limited despite the physical spectra play significant role in experimental studies \cite{Vobservation1,Vobservation2}.
So far, exact RW spectra have been calculated analytically only for the scalar nonlinear Schr\"odinger equation (NLSE) \cite{Spectra2011-1} although, numerically, spectral properties of RWs have been analysed in a couple of particular cases \cite{Zhao-2014,Akhmediev-2015}. Obtaining exact expressions for the spectra in explicit form is still problematic even for integrable models although their knowledge is important for experimental observations \cite{Vobservation1,Vobservation2} and for predictions of RWs \cite{Spectra2011-1,Spectra2011-2}.

One of the important experimental observations is the asymmetric RW spectra of the vector NLSE \cite{Vobservation1}. The latter are known as Manakov equations \cite{MM}.
Finding the RW spectra for the vector case in analytic form would fill the gap existing between the experiments and the RW theory.
We should also take into account that the Manakov equations admit a wide range of RW solutions that include new types known as nondegenerate RWs \cite{N-RW}. Each of these solutions is uniquely defined by the individual eigenvalue of the inverse scattering technique. Finding spectra of these solutions would enrich our knowledge of vector RWs.

In this work, we derived the exact analytic spectra for the RWs of Manakov equations. They cover both focusing and defocusing cases.
We have shown that in log scale, the spectra are triangular at each side of the spectrum. Such spectra are fully described by two parameters: the spectral slope and the spectral jump at zero frequency.
The latter parameter introduces asymmetry into the RW spectra.
Our theoretical results explain the the asymmetric spectra of vector RWs that have been observed in the defocusing regime \cite{Vobservation1}.

\section{exact vector RW solutions}

The Manakov equations \cite{MM} can be written in the following form:
\begin{eqnarray}
\begin{aligned}
i\frac{\partial\psi^{(1)}}{\partial t}+\frac{1}{2}\frac{\partial^2\psi^{(1)}}{\partial x^2}+\sigma(|\psi^{(1)}|^2+|\psi^{(2)}|^2)\psi^{(1)}&=0,\\
i\frac{\partial\psi^{(2)}}{\partial t}+\frac{1}{2}\frac{\partial^2\psi^{(2)}}{\partial x^2}+\sigma(|\psi^{(1)}|^2+|\psi^{(2)}|^2)\psi^{(2)}&=0.\label{eq1}
\end{aligned}
\end{eqnarray}
where $\psi^{(1)}(t,x)$, $\psi^{(2)}(t,x)$ are the two nonlinearly coupled components of the vector wave field. Parameter $\sigma=\pm1$ denotes the strength of the nonlinearity.
In the case $\sigma=1$, Eqs. (\ref{eq1}) describe the focusing
(or anomalous dispersion) regime while the case $\sigma=-1$
is related to the defocusing (or normal dispersion)
regime. These equations play a pivotal role in modelling variety of nonlinear wave phenomena in Bose-Einstein condensates \cite{BEC}, in optics \cite{OF}, and in hydrodynamics \cite{F}. The physical meaning of independent variables $x$ and $t$ depends on a particular physical problem of interest. In optics, $t$ is commonly a normalised distance along the fibre while $x$ is the normalised time in a frame moving with group velocity \cite{OF}. In the case of Bose-Einstein condensates, $t$ is time while $x$ is the spatial coordinate \cite{BEC}.

Using the Darboux transformation method \cite{DT1991}, a family of fundamental (first-order) RW solutions in concise form are given by \cite{VMRW3}
\begin{eqnarray}
\psi^{(j)}=
\psi_{0}^{(j)}\left[1+\frac{2i(\chi_{r}+b_{j})(x+\chi_{r}t)-2i\chi_{i}^2t-1}
{\gamma_{j}[(x+\chi_{r}t)^2+\chi_{i}^2t^2+1/(2\chi_{i})^2]}\right],\label{eqrw}
\end{eqnarray}
where $\psi_{0}^{(j)}$ is the vector plane wave:
\begin{equation}
\psi_{0}^{(j)}= a_j \exp{[i({\beta_j}x + \kappa_j t)]}.\label{eqpw}
\end{equation}
Here $\kappa_j=a_1^2+a_2^2- 1/2\beta_j^2$
with the amplitudes $a_j$ and the wavenumbers $\beta_j$, respectively. Moreover,
$\gamma_{j}=({\chi}_{r}+\beta_{j})^2+{\chi}_{i}^2$.
The values, ${\chi}_r\equiv\textrm{Re}[{\chi}]$ and ${\chi}_i\equiv\textrm{Im}[{\chi}]$ are the real and imaginary parts of the eigenvalue $\chi$. The latter is given by the following equation:
\begin{eqnarray}\label{eqev}
1+\sigma\sum^{2}_{j=1}\frac{a_{j}^{2}}{(\chi+\beta_{j})^{2}}=0.
\end{eqnarray}
Solution (\ref{eqrw}) depends on the background amplitudes $a_j$, background wavenumbers $\beta_j$, and the eigenvalue $\chi$.
From a physical viewpoint, the relative wavenumber between two wave components plays a key role in vector RW formation \cite{Zhao-2012,Zhao-2013,Baronio-2014}.
Without losing generality, we set ${\beta_1}=-{\beta_2}=\beta$. For simplicity, we take $a_1=a_2=a$.
Under this restriction, the solution (\ref{eqrw}) is symmetric relative to the sign change of $\beta$ and simultaneous change of the wave component.
Namely,
\begin{eqnarray}
\psi^{(2)}(\beta)=\psi^{(1)}(-\beta).
\end{eqnarray}

It was shown in our previous work \cite{N-RW} that the eigenvalue $\chi$ plays a vital role in the possibility of vector RW formation and its wave structure. From (\ref{eqev}), the explicit expressions for $\chi$ are given by
\begin{eqnarray}\label{eqev1}
\begin{aligned}
\chi_1&=-\chi_2=\sqrt{\mu-\nu},\\
\chi_3&=-\chi_4=\sqrt{\mu+\nu}.
\end{aligned}
\end{eqnarray}
where $\mu=\beta^2-\sigma a^2$, and $\nu=a\sqrt{\sigma(\sigma a^2-4\beta^2)}$.

The focusing and defocusing cases require separate analysis. When $\sigma=-1$, two eigenvalues $\chi_{1}=\chi_{2}^*$ ($*$ denotes the complex conjugate) are purely imaginary while two others are real: $\chi_{3}=\chi_{4}^*\in \mathbb{R}$.
Thus, when $\mu<\nu$ (implying $\beta\neq0$), there is only one RW solution: $\psi^{(j)}( \chi_{1})=\psi^{(j)}(\chi_{2})$. The existence of vector RWs in the defocusing case have been found in \cite{Baronio-2014}. This feature distinguishes vector NLSE case from the scalar one. These new RWs have been observed experimentally in fiber optics \cite{Vobservation1}. Their study have been done in more detail in \cite{Qin-2023}.

The focusing case ($\sigma=+1$) is less surprising. Nevertheless, it was shown in \cite{N-RW} that there are two distinctly different regimes. They follow from Eq. (\ref{eqev1}): (i) $|\beta|\leq a/2$; (ii) $|\beta|>a/2$.

(i) When $|\beta|\leq a/2$, there is only one RW solution, $\psi^{(j)}( \chi_{1})=\psi^{(j)}(\chi_{2})$ where the eigenvalue $\chi_{1}=\chi_{2}^*$ is purely imaginary. This particular case can be regarded as the vector generalization of the scalar RW. Indeed, when $\beta=0$, solution (\ref{eqrw}) $\psi^{(j)}( \chi_{1})$ bluntly coincides with the scalar-NLSE RW solution \cite{Peregrine}.

(ii) When $|\beta|>a/2$, there are two different RWs defined by $\psi^{(j)}( \chi_{1})=\psi^{(j)}( \chi_{3})$ and $\psi^{(j)}(\chi_{2})=\psi^{(j)}(\chi_{4})$ where $\chi_{1}=\chi_{3}^*$, $\chi_{2}=\chi_{4}^*$. Namely, for any given initial background parameters ($|\beta|>a/2$), there are two different RWs.
The nonlinear superposition between these two different RWs produces the nondegenerate second-order RWs,
$\psi^{(j)}( \chi_{1},\chi_{2})$ \cite{N-RW}.

\section{exact spectra of vector RWs} \label{Sec3}

Each of the two wave components $\psi^{(j)}(x,t)$ consists of an infinite number of spectral harmonics:
\begin{eqnarray}
\psi^{(j)}_{x,t}=\int_{-\infty}^{+\infty}\mathcal{F}^{(j)}_{\omega,t}~e^{i\omega x} d\omega, \label{eqj0}
\end{eqnarray}
where the Fourier components $\mathcal{F}^{(j)}_{\omega,t}$ are given by
\begin{eqnarray}
\mathcal{F}^{(j)}_{\omega,t}=\frac{1}{2\pi}\int_{-\infty}^{+\infty}{\psi^{(j)}_{x,t}}~e^{-i\omega x}dx.\label{eqj1}
\end{eqnarray}
To simplify the calculations, Eq. (\ref{eqj0}) can be rewritten as
\begin{eqnarray}
\psi^{(j)}_{x,t}=\int_{-\infty}^{+\infty}\widetilde{\mathcal{F}}^{(j)}_{\omega,t}~\psi_{0}^{(j)}~e^{i\omega x} d\omega.\label{eqj0x}
\end{eqnarray}
Here $\widetilde{\mathcal{F}}^{(j)}_{\omega,t}$ are the normalized spectra given by
\begin{eqnarray}
\widetilde{\mathcal{F}}^{(j)}_{\omega,t}=\frac{1}{2\pi}\int_{-\infty}^{+\infty}{\psi^{(j)}}/{\psi_{0}^{(j)}}~e^{-i\omega x}dx.\label{eqj1x}
\end{eqnarray}
Inserting Eq. (\ref{eqpw}) into (\ref{eqj0x}), we obtain
\begin{eqnarray}
\psi^{(j)}_{x,t}=\int_{-\infty}^{+\infty}a_j~\widetilde{\mathcal{F}}^{(j)}_{\omega,t}~e^{i \kappa_j t}~e^{i(\omega+\beta_j) x} d\omega.\label{eqj2x}
\end{eqnarray}
Comparing Eqs. (\ref{eqj0}) and (\ref{eqj2x}), we have
\begin{eqnarray}
\mathcal{F}^{(j)}_{\omega,t}=a_j~\widetilde{\mathcal{F}}^{(j)}_{\omega-\beta_j,t}~e^{i \kappa_j t}.\label{eqgp}
\end{eqnarray}
Equation (\ref{eqgp}) provides a simple transformation between the spectrum $\mathcal{F}^{(j)}_{\omega,t}$ and the normalized spectrum $\widetilde{\mathcal{F}}^{(j)}_{\omega,t}$.

Now, we perform a change of variable to the complex one in the normalized spectrum  (\ref{eqj1x}): $x\rightarrow \mathcal{Z}$, so that
\begin{eqnarray}
\widetilde{\mathcal{F}}^{(j)}_{\omega,t}=\frac{1}{2\pi}\int_{-\infty}^{+\infty}(1+\mathcal{J}_{j})
e^{-i\omega \mathcal{Z}}d\mathcal{Z},\label{eqj2}
\end{eqnarray}
where
\begin{eqnarray}
\mathcal{J}_{j}=\frac{2i(\chi_{r}+b_{j})(\mathcal{Z}+\chi_{r}t)-2i\chi_{i}^2t-1}
{\gamma_{j}[(\mathcal{Z}+\chi_{r}t)^2+\chi_{i}^2t^2+1/(2\chi_{i})^2]}.\label{eqj3}
\end{eqnarray}
Taking into account that `1' and the Dirac delta function are the Fourier transform pairs (i.e., $1\leftrightarrow\delta{(\omega)}$), the nontrivial part of the integral (\ref{eqj2}) is
\begin{eqnarray}
\mathcal{I}^{(j)}=\frac{1}{2\pi}\int_{-\infty}^{+\infty}\mathcal{J}_{j}
e^{-i\omega \mathcal{Z}}d\mathcal{Z}.\label{eqI}
\end{eqnarray}
 The roots of the binomial in the denominator of $\mathcal{J}_{j}$ are
\begin{eqnarray}
{\mathcal{Z}}_{1}=\frac{i\sqrt{4\chi_{i}^4t^2+1}}{2\chi_{i}}-\chi_{r}t,\\
{\mathcal{Z}}_{2}=-\frac{i\sqrt{4\chi_{i}^4t^2+1}}{2\chi_{i}}-\chi_{r}t,\label{eq8}
\end{eqnarray}

\begin{figure*}[htb]
\centering
\includegraphics[width=150mm]{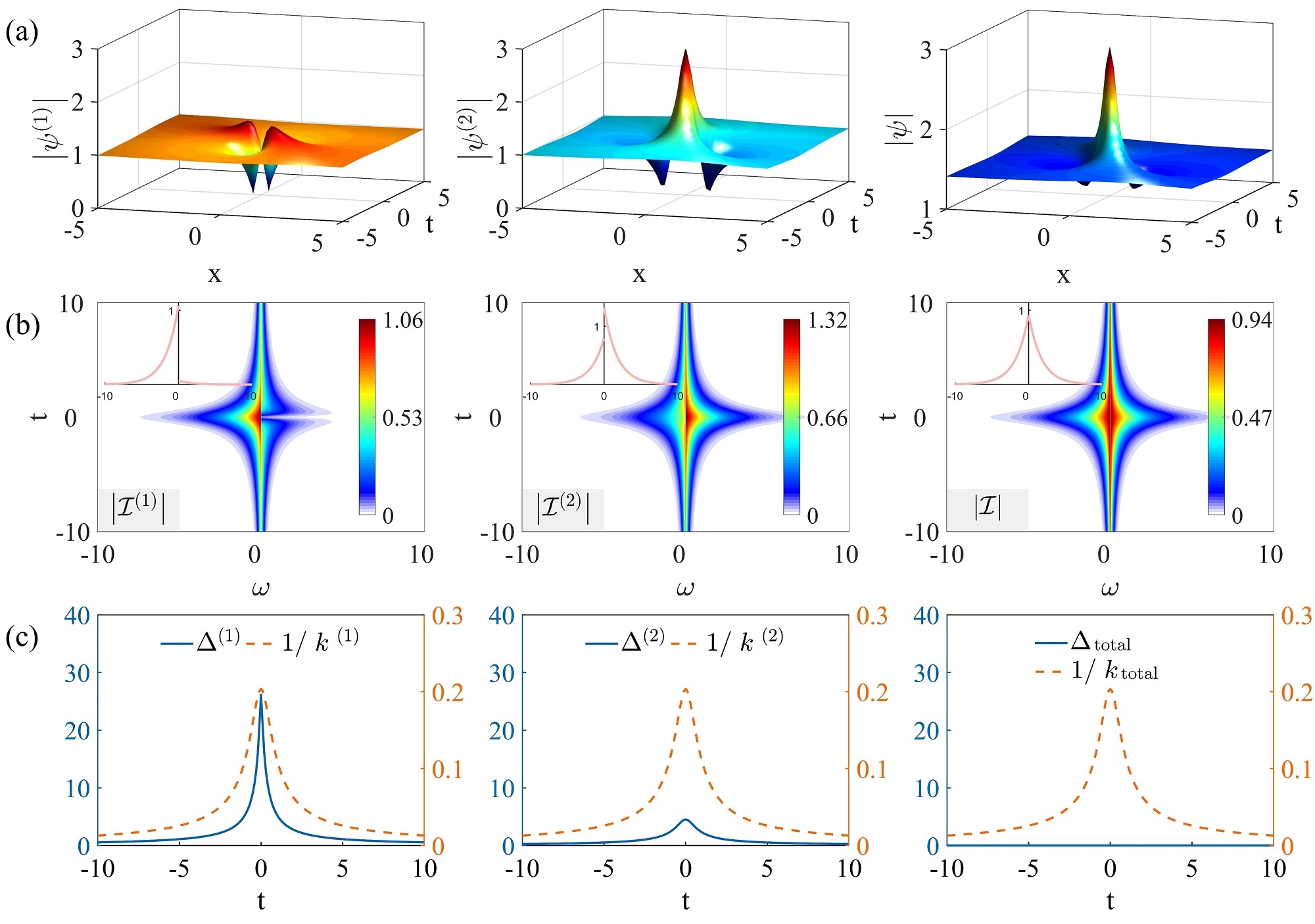}
\caption{(a) Amplitude profiles of the vector RW components $|\psi^{(1)}|$, $|\psi^{(2)}|$, given by Eqs. (\ref{eqrw}), with the eigenvalue $\chi=\chi_1$, and the total amplitude $|\psi|=\sqrt{(|\psi^{(1)}|^2+|\psi^{(2)}|^2)}$. (b) The corresponding spectra $|\mathcal{I}^{(1)}|$, $|\mathcal{I}^{(2)}|$, and $|\mathcal{I}|=\sqrt{(|\mathcal{I}^{(1)}|^2+|\mathcal{I}^{(2)}|^2)/2}$, given by Eqs. (\ref{eqzh}). The insets show the spectra at $t=0$.
(c) Evolution of the two spectral characteristics of the RWs in $t$: (1) the spectral jump $\Delta^{(j)}$ at zero frequency given by Eq. (\ref{eqDD1}) (blue solid curves), (2) the spectral slopes, $k^{(j)}=k$, given by  Eq. (\ref{eqKK1}) (orange dashed curves).
The parameters $a=1$, and $\beta=0.6$.}  \label{f1}
\end{figure*}

The roots ${Z}_{1}$ and ${Z}_{2}$ are the first-order singularities for $\mathcal{J}_j$.
Then, the residue $\mathcal{R}^{j}_{\mathcal{Z}}$ can be calculated by using the formula:
\begin{eqnarray}
\mathcal{R}^{(j)}_{\mathcal{Z}}=\frac{2i(\chi_{r}+\beta_{j})(\mathcal{Z}+\chi_{r}t)-2i\chi_{i}^2t-1}
{2\gamma_{j}(\mathcal{Z}+\chi_{r}t)}.\label{eqR}
\end{eqnarray}
Before using the residue theorem, let us pay more attention to the  imaginary parts of ${\mathcal{Z}}_{1}$ and ${\mathcal{Z}}_{2}$. They are of opposite signs (i.e., ${\textmd{Im}}[\mathcal{Z}_{1}]=-{\textmd{Im}}[\mathcal{Z}_{2}]$).
This makes the calculations more difficult.
Since the integral (\ref{eqI}) depends on the imaginary parts of ${\mathcal{Z}}_{1,2}$, we shall calculate it on a case-by-case basis.
Namely, substituting ${\mathcal{Z}}_{1,2}$ into Eq.
(\ref{eqR}) and applying the residue theorem yields
\begin{eqnarray}
\mathcal{I}^{(j)}
&=&\left\{
\begin{aligned}
&i\mathcal{R}^{j}_{\mathcal{Z}_1}e^{-i\omega \mathcal{Z}_1},&  &{\textmd{sgn}(\omega)=-\textmd{sgn}(\chi_i),}\\
&i\mathcal{R}^{j}_{\mathcal{Z}_2}e^{-i\omega \mathcal{Z}_2},&  &{\textmd{sgn}(\omega)=\textmd{sgn}(\chi_i).}\label{eq13}
\end{aligned}
\right.
\end{eqnarray}
Collecting the results above, we have
\begin{eqnarray}
\widetilde{\mathcal{F}}^{(j)}=\mathcal{I}^{(j)}+\delta(\omega).
\end{eqnarray}
The Delta delta function $\delta(\omega)$ appears due to the infinite size of the background. Removing it does not affect the spectra.
Using the relation $\textmd{sgn}(\omega)\textmd{sgn}(\chi_{i})=\textmd{sgn}(\omega \chi_{i})$,
we obtain the spectra of the vector RW:
\begin{eqnarray} \label{eqzh}
\mathcal{I}^{(j)}=&&
\left[-\frac{\chi_{r}+\beta_{j}}{\gamma_{j}}+\textmd{sgn}(\omega \chi_{i})\frac{2i\chi_{i}^{3}t+\chi_{i}}{\gamma_{j}\sqrt{4\chi_{i}^4t^2+1}}\right]  \\ \nonumber
&&\times \exp\left[{-\textmd{sgn}(\omega\chi_{i})\frac{\omega\sqrt{4\chi_{i}^4t^2+1}}{2\chi_{i}}+i\chi_{r}\omega t}\right].
\end{eqnarray}
After applying $\textmd{sgn}(\chi_i)\chi_i=|\chi_i|$, the modulus of Eq. (\ref{eqzh}) becomes:
\begin{eqnarray}
|\mathcal{I}^{(j)}|
=\left\{
\begin{aligned}
|\Delta_{0}^{(j)}-\Delta_{s}^{(j)}(t)| \exp\left[\omega S(t)\right],~~~\omega<0,\\
|\Delta_{0}^{(j)}+\Delta_{s}^{(j)}(t)|\exp\left[-\omega S(t)\right],~ \omega>0,\label{eqIM}
\end{aligned}
\right.
\end{eqnarray}
where
\begin{eqnarray}
&&\Delta_{0}^{(j)}=-\frac{\chi_{r}+\beta_{j}}{\gamma_{j}},~~~\Delta_{s}^{(j)}(t)=\frac{2i|\chi_{i}|^{3}t+|\chi_{i}|} {\gamma_{j}\sqrt{4\chi_{i}^4t^2+1}},~~~~~~~~~\label{eqd0d}\\
&&S(t)=\frac{\sqrt{4\chi_{i}^4t^2+1}}{2|\chi_{i}|}.\label{eqst}
\end{eqnarray}
Eqs. (\ref{eqIM}) is the result that we were looking for. However, it still need more detailed analysis.

One immediate conclusion that follows from (\ref{eqIM}) is that once $\Delta_0^{(j)}\neq0$ (implying $\beta\neq0$), the spectra of the vector RWs at any given $t$ are asymmetric relative to the sign change of the frequency. Indeed, $\mathcal{I}^{(j)}_{\omega_-}\neq\mathcal{I}^{(j)}_{\omega_+}$.
The nonzero relative wavenumber ($\beta\neq0$) of the two components leads to the spectral asymmetry of the vector field.
This feature distinguishes the vector RWs from the scalar ones that have a symmetric spectra \cite{Spectra2011-1}.

Alternatively, when $\beta=0$ (implying $\Delta_0^{(j)}=0$), the Manakov equations in the focusing regime ($\sigma=1$) are reduced to the scalar NLSE. Correspondingly, the spectra of vector RWs (\ref{eqIM}) are becoming symmetric:
\begin{eqnarray}
|\mathcal{I}^{(j)}|=\frac{1}{\sqrt{2}a } \exp\left[{-|\omega|\frac{\sqrt{16a^4t^2+1}}{2\sqrt{2}a}}\right].\label{eqbb}
\end{eqnarray}
This is exactly the spectrum obtained in Ref. \cite{Spectra2011-1} for the RW of the scalar NLSE.

Figure \ref{f1} illustrates the evolution of RW in space and its spectrum. Figure \ref{f1}(a) shows the amplitude profiles of the vector RW with $\chi=\chi_1$ and $\beta\neq0$. The first component $|\psi^{(1)}|$ has a four-petal structure while the second component $|\psi^{(2)}|$ exhibits a bright peak above the background.
The total amplitude $\sqrt{(|\psi^{(1)}|^2+|\psi^{(2)}|^2)}$ also displays a bright peak due to choice of the focusing coefficient in the Manakov equations.

Figure \ref{f1}(b) shows the evolution of the corresponding spectra.
For each wave component, we observe spectral expansion at negative $t$ and spectral contraction at positive $t$. The spectra of the two components are asymmetric relative to the sign change of the frequency $\omega$. The spectrum of first wave component $|\psi^{(j)}|$ experiences the extreme spectral asymmetry at $t=0$.
This is related to the similar asymmetry in the spectra of vector Akhmediev breathers
\cite{VAB2021,VAB2022,VAB-Df2022} although the spectra in the latter case are discrete.

The spectrum of the total wave field $|\mathcal{I}|=\sqrt{(|\mathcal{I}^{(1)}|^2+|\mathcal{I}^{(2)}|^2)/2}$ in Fig. \ref{f1}(b) is symmetric around $\omega=0$. This follows directly from Eq.(\ref{eqIM}):
\begin{eqnarray}\label{eqxd}
|\mathcal{I}^{(1)}_{\omega_-}|^{2}+|\mathcal{I}^{(2)}_{\omega_-}|^{2}=
|\mathcal{I}^{(1)}_{\omega_+}|^{2}+|\mathcal{I}^{(2)}_{\omega_+}|^{2}.
\end{eqnarray}

\section{Two spectral characteristics of the RW}\label{Sec6a}

Let us turn to the spectral characteristics of the vector RWs. The spectra in experiments are usually measured in log scale (in decibels).
Using the common definition $\left|\mathcal{I}^{(j)}\right|^2_{dB}=20 \lg{\left|\mathcal{I}^{(j)}\right|}$, we obtain, from (\ref{eqIM}),
\begin{eqnarray}
|\mathcal{I}^{(j)}|^2_{dB}
=\left\{
\begin{aligned}
20 \lg |\Delta_{0}^{(j)}-\Delta_{s}^{(j)}(t)| +\frac{20~S(t)}{\lg10}\omega,~\omega<0,\\
20 \lg |\Delta_{0}^{(j)}+\Delta_{s}^{(j)}(t)| -\frac{20~S(t)}{\lg10}\omega,~ \omega>0.\label{eqidb}
\end{aligned}
\right.
\end{eqnarray}
The function $\left|\mathcal{I}^{(j)}\right|^2_{dB}$ in (\ref{eqidb}) is a linear function of $\omega$ both in the negative and positive frequency regions.
Two distinctive parameters can be defined to characterise these spectra.

\subsection{The slopes of the spectra}\label{Sec6.1b}
The absolute values of the slopes of the spectra of vector RWs at the negative and positive frequencies $k^{(j)}_{\omega_\pm}$ are given by
\begin{eqnarray}
k^{(j)}_{\omega_\pm}&&\equiv \left| \frac{|\mathcal{I}^{(j)}|^2_{dB}({\omega_1}_\pm)-|\mathcal{I}^{(j)}|^2_{dB}({\omega_2}_\pm)}
{{\omega_1}_\pm-{\omega_2}_\pm}\right|,\label{eqKK}
\end{eqnarray}
where ${\omega_{1}}_\pm$, ${\omega_{2}}_\pm$ are two different (arbitrary) frequencies at the positive and negative regions respectively.
Substituting Eq. (\ref{eqidb}) into (\ref{eqKK}), we obtain the explicit forms of the spectral slopes:
\begin{eqnarray}
k^{(j)}_{\omega_\pm}
=\frac{20}{\lg 10}S(t)=\frac{20}{\lg 10}\frac{\sqrt{4\chi_{i}^4t^2+1}}{2|\chi_{i}|}.\label{eqKK1}
\end{eqnarray}
The coefficients $k^{(j)}_{\omega_\pm}$ in (\ref{eqKK1}) do not depend on $\omega$.
Also, the slopes of the spectra at negative and positive frequencies are equal,
\begin{eqnarray}\label{}
k^{(j)}=k^{(j)}_{\omega_+}=k^{(j)}_{\omega_-}.
\end{eqnarray}
Moreover, $k^{(j)}$ does not depend on the sign of $\beta_j$.
Then, the spectral slopes for the two wave components are also equal:
\begin{eqnarray}\label{eqk1k2}
k^{(1)}=k^{(2)}=k.
\end{eqnarray}
The reciprocal of the slope $1/k^{(j)}$ describes the degree of spectral expansion
at each stage of the evolution.
Figure \ref{f1} (c) shows the evolution of $1/k^{(j)}$ in time $t$ separately for each wave component $\psi^{(1)}$ and $\psi^{(2)}$ and for the total wave field (orange dashed lines).
At $t=0$, the value of $1/k$ reaches its maximum
\begin{eqnarray}
\frac{1}{k}= \frac{\lg{10}~|\chi_i|}{10}.\label{eqkk}
\end{eqnarray}
This is the expression for the maximum broadening of the spectra that occurs at $t=0$.

\subsection{The spectral jump}\label{Sec6.2a}
Although the slopes of the spectra at the negative and positive frequencies are the same, the spectra are still asymmetric. The asymmetry comes from the spectral jumps $\Delta^{(j)}$ at zero frequency.
 The spectral jump is different for each wave component:
\begin{eqnarray}
\Delta^{(j)}\equiv\left||\mathcal{I}^{(j)}|^2_{dB}(\omega_-)- |\mathcal{I}^{(j)}|^2_{dB}(\omega_+)\right|\label{eqDD}.
\end{eqnarray}
Inserting Eq. (\ref{eqidb}) into (\ref{eqDD}), we obtain
\begin{eqnarray}
\Delta^{(j)}=20 \left|\lg {\frac{|\Delta_0^{(j)}+\Delta_S^{(j)}(t)|}{|\Delta_0^{(j)}-\Delta_S^{(j)}(t)|}}\right|.\label{eqDD1}
\end{eqnarray}
Generally, the jump is nonzero $\Delta^{(j)}\neq0$, as soon as $\beta\neq0$.

\begin{figure}[htb]
\centering
\includegraphics[width=88mm]{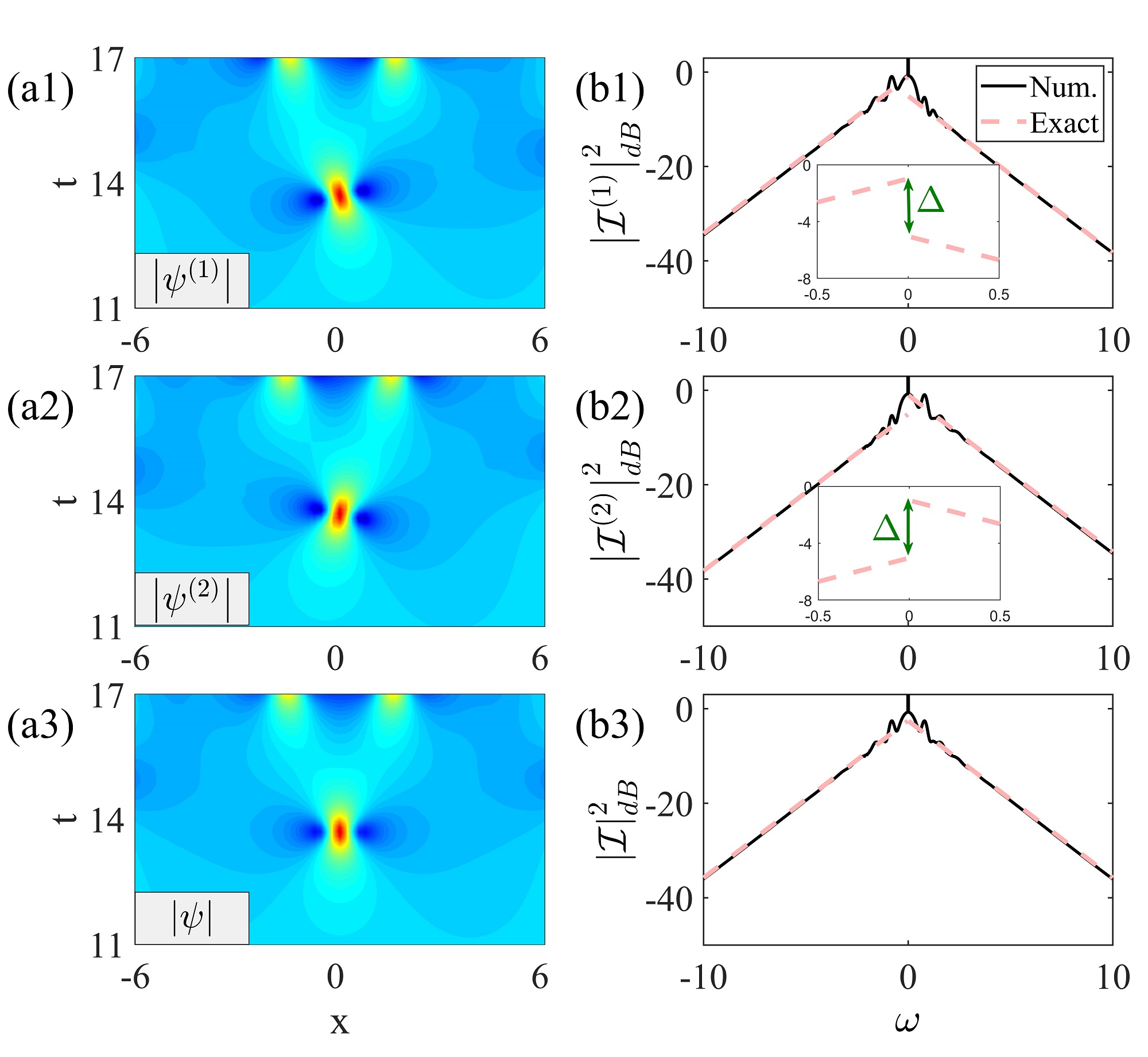}
\caption{(a1-a3) Amplitude profiles of a single RW obtained in \ numerical simulations with the initial conditions (\ref{eqic}) for parameters $\beta=0.3$ and $a=1$.
(b1-b3) RW spectra calculated at the time of the maximum broadening ($t=13.7$ in (a)) (black solid lines) overlapped with the spectra given by the exact
solutions (\ref{eqidb}) for $\chi=\chi_1$ at $t=0$ (pink dashed lines). The insets show the enlarged spectra at the top of the curves.
}\label{f2}
\end{figure}

Let us consider separately the cases $\sigma=-1$ and $\sigma=1$.
When $\sigma=-1$, we have $\chi_r=0$. Then, it follows, from (\ref{eqDD1}), that
the spectral jumps for the two wave components are equal: $\Delta^{(1)}=\Delta^{(2)}\neq0$. This result is in agreement with the experimental observation of the asymmetric spectra of vector RWs in the defocusing regime presented in Figs. 6(g) and 6(h) of Ref. \cite{Vobservation1}.
On the other hand, in the focusing case, $\sigma=1$, the two spectral jumps are equal, $\Delta^{(1)}=\Delta^{(2)}$, only if $|\beta|\leq a/2$. This can be seen from Figs. \ref{f2} (b1) and \ref{f2} (b2) showing the spectra of the two wave components.
However, if $|\beta|> a/2$, the two spectral jumps differ: $\Delta^{(1)}\neq\Delta^{(2)}$. This can be seen from the plots of $\Delta^{(j)}$ shown in Fig. \ref{f1} (c).
The spectral jumps $\Delta^{(j)}$ also reach their maxima at $t=0$.
These values are given by
\begin{eqnarray}
\Delta^{(j)}(t=0)=20 \lg{\left|\frac{|\chi_{i}|+|(\chi_{r}+\beta_{j})|}{|\chi_{i}|-|(\chi_{r}+\beta_{j})|}\right|}. \label{eqdd}
\end{eqnarray}

\section{Numerical simulations}\label{Sec6b}

We simulated numerically the vector RWs in order to verify our theoretical findings.
One possible way of simulations would be taking the exact solution (\ref{eqrw}) at any negative $t$ as the initial condition. Naturally, these simulations will result in the same profiles of the RWs as given by Eq. (\ref{eqrw}). Another way to excite the RW is to use the initial condition in the form of a plane wave with a small localised perturbation given by a simple known function. The advantage of using such initial conditions is that these simulations will demonstrate the robustness of the RWs.
Consequently, these simulations would be easier to implement in future experiments.

In our simulations, we have solved the Manakov equations numerically using the split-step Fourier method. As the initial condition, we used the following function: \begin{equation}\label{eqic}
\psi^{(j)}=\psi_{0}^{(j)}\left[1+\epsilon L_p(x/x_0)\right],
\end{equation}
where $L_p(x/x_0)$ is either a sech-function $L_p=\textmd{sech}(x/x_0)$ or a Gaussian $L_p=\exp{(-x^2/x_0^2)}$. In each case, $x_0$ is the width of the localised perturbation while $\varepsilon$ is the small amplitude of perturbation ($\varepsilon\ll1$). In all simulations, we selected the Gaussian perturbation with $\epsilon=0.01$ and $x_0=6$. We also restricted ourselves to the focusing regime ($\sigma=1$).
This allowed us to reveal two specific RW dynamics for $|\beta|\leq a/2$ and $|\beta|>a/2$, discussed above and in our previous work \cite{N-RW}.

\subsection{Excitation of a single RW for $|\beta|\leq a/2$}\label{Sec6.1a}

Figure \ref{f2} shows the results of numerical simulations of the process of a single RW excitation starting with the initial condition (\ref{eqic}) for $|\beta|\leq a/2$ and $\sigma=1$.
Figures \ref{f2}(a1) and \ref{f2}(a2) display the amplitude profiles of the two vector field components $|\psi^{(1)}|$ and $|\psi^{(2)}|$, respectively. The total amplitude profile $|\psi|$ is shown in Fig \ref{f2}(a3). As we can see, a single RW is excited with high accuracy despite the approximate initial conditions are used.
Using the data in Fig \ref{f2}(a), we calculated the corresponding spectra at the time of maximum broadening ($t=13.7$). These are presented in Figs. \ref{f2}(b1-b3) by black solid lines.
The spectra given by the exact solution (\ref{eqidb}) at $t=0$ are also shown by pink dashed lines. The spectra found from the numerical simulations are in good agreement with the exact results.
In particular, we found that the slopes of the spectral curves
are equal ($k^{(1)}=k^{(2)}=k$) in agreement with Eq. (\ref{eqk1k2}).
The minimal slopes of the spectra in Figs. \ref{f2}(b1) and \ref{f2}(b2) obtained numerically are in good agreement with  the exact result given by Eq. (\ref{eqkk}).
The spectral jumps at zero frequency of the vector wave field components that are shown in Figs.\ref{f2}(b1) and \ref{f2}(b2) are also equal, i.e.,  $\Delta^{(1)}=\Delta^{(2)}$, as predicted by Eq. (\ref{eqdd}). The scale is magnified in the insets of these figures for better visibility.

\begin{figure*}[htb!]
\centering
\includegraphics[width=150mm]{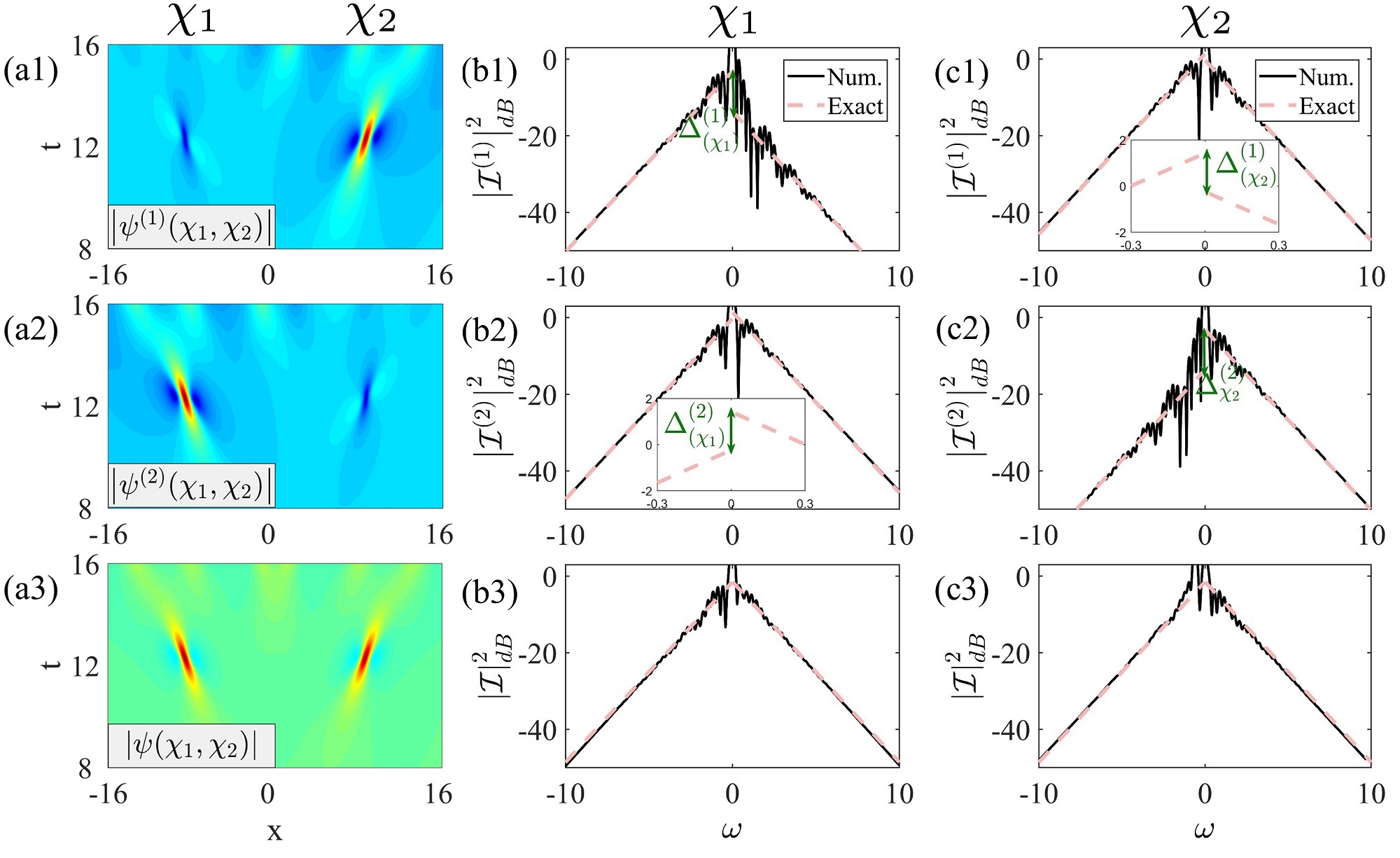}
\caption{(a) Amplitude profiles of the nondegenerate second-order RWs $\psi^{(j)}(\chi_1,\chi_2)$ obtained in numerical simulations with the initial conditions (\ref{eqic}) with $\beta=0.9$ and $a=1$.
(b) Spectra of a single (l.h.s.) RW $\psi^{(j)}( \chi_{1})$  calculated in numerical simulations at the position of its maximum amplitude (black solid lines) overlapped with the exact analytical results described by Eqs. (\ref{eqidb}) at $t=0$ with $\chi=\chi_1$ (pink dashed lines). (c) The same as in (b) but for the second (r.h.s.) RW with $\chi=\chi_2$.} \label{f3}
\end{figure*}

\subsection{Excitation of nondegenerate RW with $|\beta|>a/2$}\label{Sec6.2b}

Figure \ref{f3}(a) shows the results of numerical simulations for $|\beta|>a/2$ and $\sigma=1$ started from the initial conditions (\ref{eqic}).
In sharp contrast to the previous case, here, the initial perturbation develops into two different fundamental RWs which are well-separated in $x$. These are dark and `bright' RWs which
are the nondegenerate RWs formed by the nonlinear superposition of two RWs with different eigenvalues, i.e.,  $\psi^{(j)}(\chi_1)$ and $\psi^{(j)}(\chi_2)$.

Figures \ref{f3}(b) and \ref{f3}(c) show in black solid lines the spectra of these two fundamental RWs at the time of the maximal broadening ($t=13.7$). They are overlapped with the corresponding spectra at $t=0$ obtained from the exact solutions (\ref{eqidb}) shown in pink dashed lines.
The spectra obtained in numerical simulations for each fundamental RW are obtained using the direct Fourier transform of the wave field multiplied by a super-Gaussian function \cite{Spectra2011-2}.
Again, the spectra obtained in numerical simulations are in good agreement with the exact results.
Despite the fundamental RWs correspond to two different eigenvalues $\chi_1=-\chi_2$, their spectral slopes are identical, $k_{(\chi_1)}=k_{(\chi_2)}$. This can be seen from the comparison of the spectral profiles shown in Figs. \ref{f3}(b) and \ref{f3}(c). This result is also in agreement with Eq. (\ref{eqKK1}).

The spectra of the two fundamental RWs have spectral jumps, $\Delta^{(j)}_{(\chi_1)}$ and $\Delta^{(j)}_{(\chi_2)}$, at zero frequency. This can be seen in Figs. \ref{f3} (b) and \ref{f3} (c).
One immediate noticeable result is that the spectral jumps for the two field components are different, $\Delta^{(j)}_{(\chi_1)}\neq\Delta^{(j)}_{(\chi_2)}$.
However, the values of $\Delta^{(j)}$ are symmetric relative to the sign change of the eigenvalue $\chi$ and simultaneous change of the wave components. Namely,
\begin{eqnarray}
\Delta^{(1)}_{(\chi_1)}&=&\Delta^{(2)}_{(-\chi_1)}\equiv \Delta^{(2)}_{(\chi_2)}, \\
\Delta^{(2)}_{(\chi_1)}&=&\Delta^{(1)}_{(-\chi_1)}\equiv \Delta^{(1)}_{(\chi_2)}.
\end{eqnarray}
These relations can be verified using Eq. (\ref{eqDD1}) with $\chi_1=-\chi_2$.

\begin{figure*}[htb!]
\centering
\includegraphics[width=150mm]{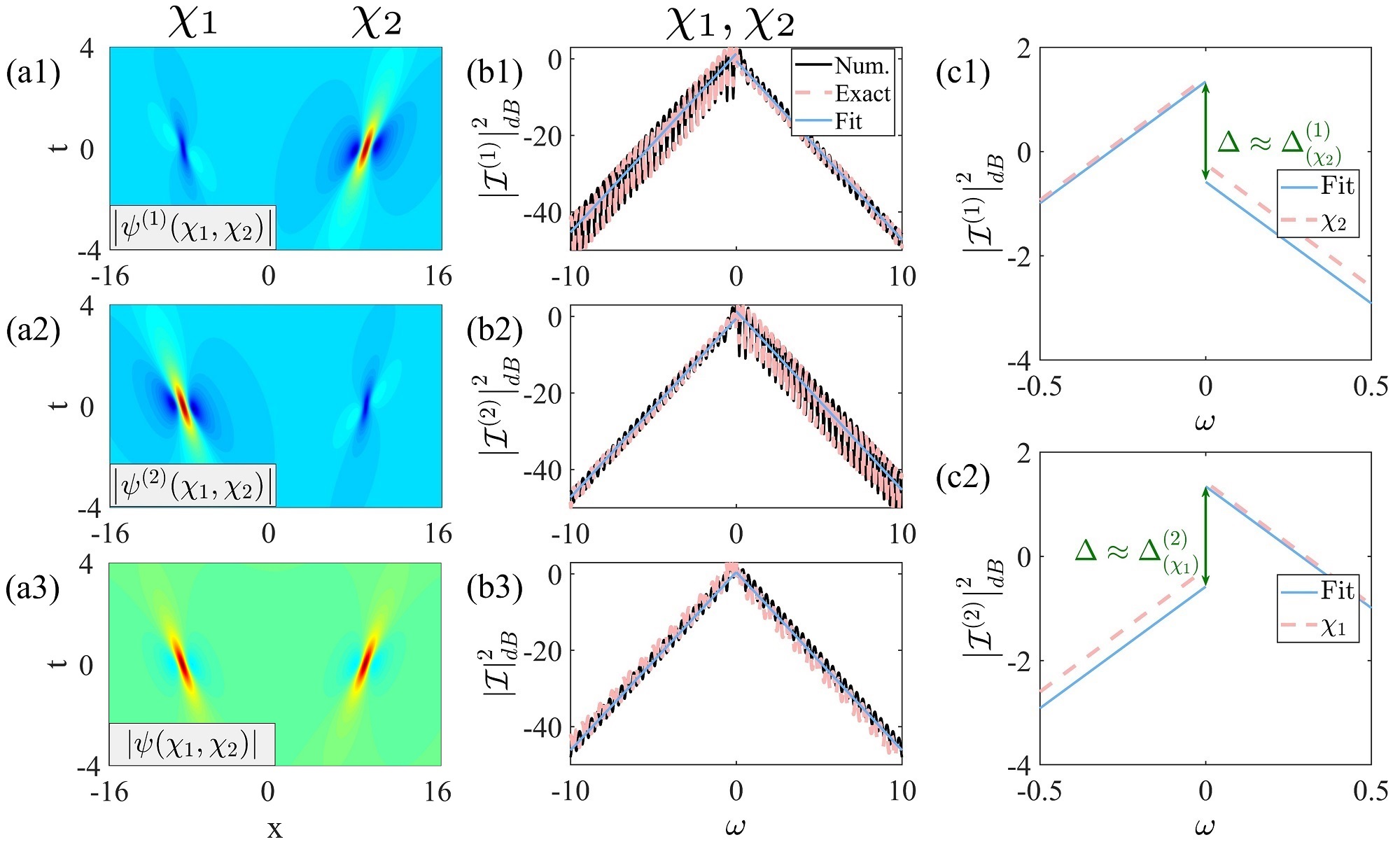}
\caption{(a) Amplitude profiles of nondegenerate RWs given by the exact solution $\psi^{(j)}(\chi_1,\chi_2)$.
(b) Spectra of this exact solution at $t=0$ (pink dashed curves) overlapped with the spectra obtained from the numerical simulations shown in Fig. \ref{f3}(a) at $t=13.7$ (black solid curves).  The fitting straight lines with the same slope as the spectra are shows in blue colour.
(c1) and (c2) Enlarged parts of (b1) and (b2) at the top of the spectra, respectively.
Parameters of the exact solution are $a=1$, $\beta=0.9$, $x_1=-8.727$, $x_2=8.727$, $t_1=0.7$, $t_2=-1.45$.}  \label{f4}
\end{figure*}

\begin{figure}[htb!]
\centering
\includegraphics[width=86mm]{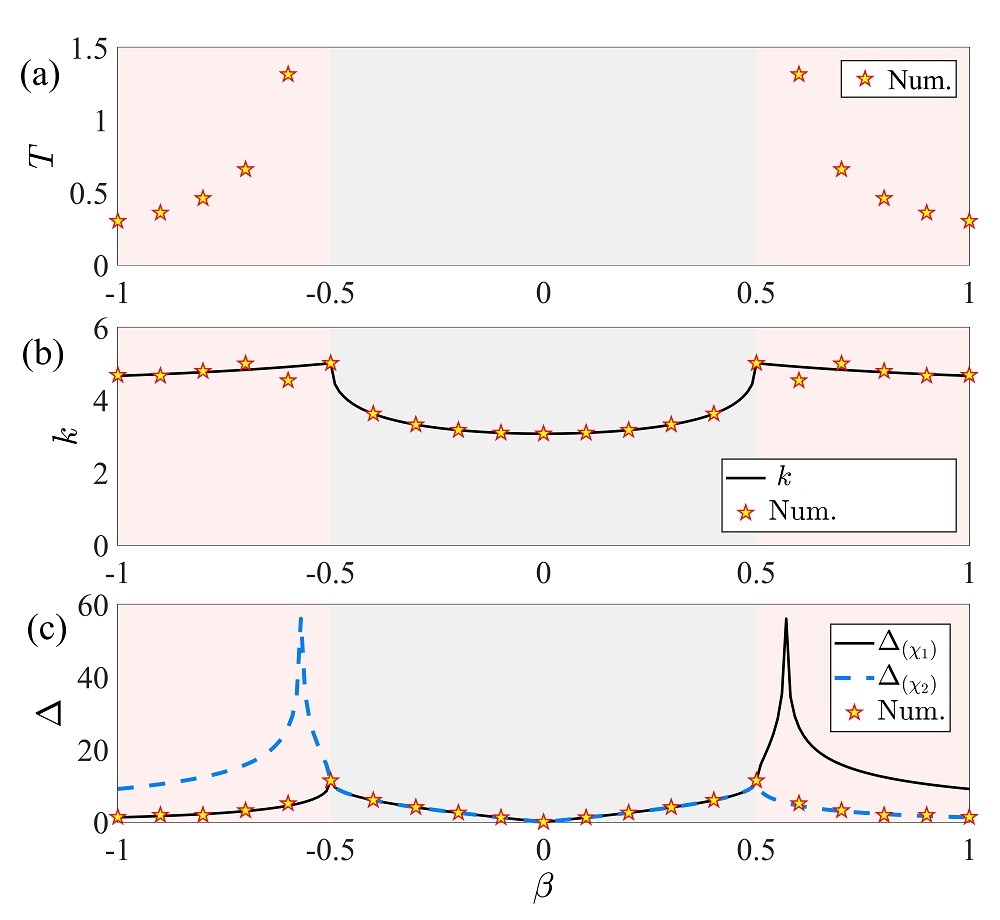}
\caption{(c) Oscillation period $T$ of non-degenerate RW spectra in Fig.\ref{f4}(b) versus $\beta$.
(b) The spectral slope $k$ versus $\beta$.  Black solid line shows the results obtained from Eq. (\ref{eqkk}) while red stars present the results obtained from numerical simulations.
(b) Spectral jumps at zero frequency versus $\beta$. Black solid lines show the results obtained from Eq. (\ref{eqDD1}) with $\chi=\chi_1$ while blue dashed lines correspond to Eq. (\ref{eqDD1}) with $\chi=\chi_2$. Red stars show the results obtained from numerical simulations.}  \label{f5}
\end{figure}

The exact solution that corresponds to the above simulations contains two nondegenerate RWs. It is the second order solution $\psi^{(j)}(\chi_1,\chi_2)$ of Manakov equations. It describes the nonlinear superposition of two fundamental RWs, $\psi^{(j)}(\chi_1)$ and $\psi^{(j)}(\chi_2)$.
The details are given in Appendix.
The solution involves two eigenvalues $\{\chi_1, \chi_2\}$ and four free parameters $\{x_1, x_2, t_1, t_2\}$ describing the positions of the RWs in $x$ and $t$. The amplitude profiles of this solution are shown in Fig. \ref{f4}(a).
As we can see, the exact solution $\psi^{(j)}(\chi_1, \chi_2)$ with an appropriate choice of the above parameters is the same as the one found in numerical simulations presented in Fig. \ref{f3}(a).

The spectra of the second-order solution $\psi^{(j)}(\chi_1,\chi_2)$ can be obtained by performing its direct Fourier transform.
Figures \ref{f4}(b) show the spectral profiles of the exact RW solution at $t=0$ (pink dashed lines) along with the spectra of the solution found in numerical simulations presented in Fig. \ref{f3}(a) at $t=13.7$ (black solid lines). There is good  agreement between the two spectra. In particular, the spectra exhibit oscillations due to the nonlinear interaction of the two nondegenerate RWs. The oscillation period decreases as $\beta$ increases. This can be seen from Fig. \ref{f5}(a).

Except for the oscillations, the spectra of nondegenerate RWs still exhibit a triangular shape. The straight lines fitting the wings in Fig. \ref{f4}(b) are shown in blue colour.
The slopes of the spectra $k_{(\chi_1, \chi_2)}$ are the same as for the fundamental RW spectra given by Eq. (\ref{eqk1k2}), i.e., $k_{(\chi_1, \chi_2)}=k$. This is illustrated in Fig. \ref{f4}(c).
In the cases of a single and the nondegenerate RW excitations, the slopes of the spectra can be described exactly by Eq. (\ref{eqKK1}). This is illustrated in Fig. \ref{f5}(b).

The asymmetry of the spectra of nondegenerate RWs at positive and negative frequencies is caused by the presence of spectral jumps $\Delta^{(j)}_{(\chi_1,\chi_2)}$ at zero frequency.
The spectral jumps of the two field components are equal $\Delta=\Delta^{(j)}_{(\chi_1,\chi_2)}$ but have opposite directions as can be seen from Fig.  \ref{f4}(c).
Remarkably, the spectral jump for the nondegenerate RWs can be described by the exact result for the fundamental RW given by Eq. (\ref{eqDD1}). This is illustrated  in Fig.  \ref{f4}(c). The spectral jump of the nondegenerate RWs is given approximately by the smaller spectral jump of the two fundamental RWs. Namely,
\begin{equation}
\Delta\approx\min\left\{\Delta^{(j)}_{(\chi_1)},\Delta^{(j)}_{(\chi_2)}\right\}.\label{eqmin}
\end{equation}
Figure \ref{f5}(c) shows the dependance of the spectral jumps of RWs on $\beta$. The results of numerical simulations are shown by the red stars while the exact analytic solution according to (\ref{eqDD1}) is shown by the blue solid curve.  In the region $|\beta|\leq a/2$, we have $\Delta^{(j)}_{(\chi_1)}=\Delta^{(j)}_{(\chi_2)}$ while
in the region $|\beta|>a/2$, the results of numerical simulations always fall on the lower curve with smaller spectral jump among the two calculated for the two fundamental RWs.
This result ones again confirms the validity of Eq. (\ref{eqmin}).

\section{Conclusions}\label{Sec5}

In conclusion, we have studied analytically and numerically the spectral properties of vector RWs for Manakov equations.
Based on the exact analytic spectra of RWs, showing triangular shape, we introduced  two characteristic quantities, namely, the slope of the spectra $k$ and the spectral jump $\Delta^{(j)}$  at zero frequency. Such quantities can be used not only to characterise the spectral properties of a single RW but also the spectra of complex nondegenerate RW superpositions.
Our work provides a solid theoretical foundation for future experimental observations of the RW spectra.

\section*{ACKNOWLEDGEMENTS}
The work is supported by the NSFC (Grants No. 12175178, and No. 12047502),
the Natural Science basic Research Program of Shaanxi Province (Grant No. 2022KJXX-71), and Shaanxi Fundamental
Science Research Project for Mathematics and Physics (Grant No. 22JSY016).

\begin{appendix}
\section{Exact RW solutions}
The exact fundamental and second-order solutions solutions describing the single and the nondegenerate RWs of the Manakov equations are constructed by the Darboux transformation \cite{DT1991}.
The equivalent linear system (Lax pair) of vector NLSEs (\ref{eq1}) is given by
\begin{eqnarray}
\Psi_x=U\Psi,~~\Psi_t=V\Psi,\label{lax}
\end{eqnarray}
with
\begin{eqnarray}
U=i\left[\frac{\lambda}{2}(S+I)+Q\right],~~~~~~~~~~~~~~~~\nonumber\\
V=i\left[\frac{\lambda^2}{4}(S+I)+\frac{\lambda}{2}Q-\frac{1}{2}S(Q^2+iQ_x)+B\right],\nonumber
\end{eqnarray}
where
\begin{eqnarray}
Q=\begin{pmatrix} 0 & \sigma{\psi_{0}^\dag}  \\ {\psi_{0}} & 0 \end{pmatrix},
~~S=\textmd{diag}\{1,-1,-1\}.
\nonumber
\end{eqnarray}
Here, ${\psi_{0}}=(\psi_{0}^{(1)}, \psi_{0}^{(2)})$ and $\dag$ represents the Hermite conjugation.
Moreover, $\lambda$ denotes the spectral parameter, $I$ is an identity matrix, and $B=(a_1^2+a_2^2)I$.

The vector eigenfunctions of the linear system (\ref{lax}) can be calculated as
\begin{eqnarray}
\Psi_m=\begin{pmatrix} \Psi_{0m}+\tilde{\Psi}_{0m}  \\
\frac{\psi_{0}^{(1)}{\Psi}_{0m}}{(\beta_1+{\chi}_{[m]})}+\frac{\psi_{0}^{(1)}\tilde{\Psi}_{0m}(\chi_{[m]}+\beta_1-1)}{(\beta_1+\chi_{[m]})^2}\\
\frac{\psi_{0}^{(2)}{\Psi}_{0m}}{(\beta_2+{\chi}_{[m]})} +\frac{\psi_{0}^{(2)}\tilde{\Psi}_{0m}(\chi_{[m]}+\beta_2-1)}{(\beta_2+\chi_{[m]})^2} \\ \end{pmatrix},\label{psim}
\end{eqnarray}
where
\begin{eqnarray}
&\Psi_{0m}&=i(\bm x+\chi_m\bm t+1)\tilde{\Psi}_{0m},\nonumber\\
&\tilde{\Psi}_{0m}&=\exp\left\{i\chi_{[m]}\bm x+i\left[\frac{\chi_{[m]}^2}{2}+(1-\sigma)(a_1^2+a_2^2)\right]\bm t\right\},\nonumber
\end{eqnarray}
with $\bm x=x-x_{m}$, and $\bm t=t-t_{m}$. Here, the spectral parameter should be: $$\lambda_{[m]}=\chi_{[m]}-\sum^2_{j=1}\frac{\sigma a_j^2}{\chi_{[m]}+\beta_j},$$ with
$\chi_{[m]}$ being the eigenvalue of the linear Lax pair system, which is given by
\begin{eqnarray}
1+\sum_{j=1}^{2}\frac{\sigma a_j^2}{({\chi}_{[m]}+\beta_j)^2}=0.\label{eqchi-aj}
\end{eqnarray}

When $m=1$, we obtain the fundamental (first-order) RW solution by performing the Darboux transformation as follows:
\begin{eqnarray}
&&\psi^{(j)}[1]=\psi_{0}^{(j)}+(\lambda^*_{[1]}-\lambda_{[1]})(P[1])_{j+1, 1},\\
&&P[1]=\frac{\Psi_1\Psi^\dag_1\Xi}{\Psi^\dag_1 \Xi\Psi_1}, ~~~\Xi=\textmd{diag}\{1,\sigma,\sigma\}.
\end{eqnarray}
Here, $\ast$ represents the complex conjugation, $\Psi_1$ is the special solution (\ref{psim}) as $\chi_{[m]}=\chi_{[1]}$.
$(P[1])_{j+1, 1}$ represent the elements of the matrix $P[1]$ in the first column, ($j+1$)th row.
The simplified form of the fundamental RW solution with $t_{1}=x_{1}=0$ is given by Eq. (\ref{eqrw}).
Fundamental vector RW with $\chi_{[1]}=\chi_{1}$ given by Eq. (\ref{eqev1}) is shown in Fig. \ref{f1}.

The nondegenerate second-order RWs reported here exist only in the focusing regime ($\sigma=1$). To obtain the exact solutions, we shall perform the second step of the Darboux transformation with $m=2$.
We employ $\Psi_2$ [Solution (\ref{psim}) as $\chi_{[m]}=\chi_{[2]}$] which is mapped to
\begin{eqnarray}
&&\Psi_2[1]=T [1]|_{\lambda=\lambda_{[2]}}\Psi_2, \nonumber\\
&&T[1]=I+\frac{\lambda^*_{[1]}-\lambda_{[1]}}{\lambda-\lambda^*_{[1]}}P[1].
\end{eqnarray}
Then, the second-order RW solution can be given by
\begin{eqnarray}
&&\psi^{(j)}[2]=\psi^{(j)}[1]+(\lambda^*_{[2]}-\lambda_{[2]})(P[2])_{j+1,1},\nonumber\\
&&P[2]=\frac{\Psi_2[1]\Psi^\dag_2[1]\Xi}{\Psi^\dag_2[1]\Xi\Psi_2[1]}.
\end{eqnarray}
The nondegenerate second-order RWs depend on the two different eigenvalues ($\chi_{[1]}=\chi_{1}$, $\chi_{[2]}=\chi_{2}$) and the four free parameters $\{x_1, x_2, t_1, t_2\}$. One example of the amplitude distributions of such RWs are shown in Fig. \ref{f4}(a).
\end{appendix}

\end{CJK}


\begin{thebibliography}{99}
\bibitem{RW09} N. Akhmediev, A. Ankiewicz, and M. Taki, Waves that appear from nowhere and disappear without a trace, \href{https://www.sciencedirect.com/science/article/pii/S0375960108017945}{Phys. Lett. A \textbf{373}, 675 (2009)}.
\bibitem{Shrira10} V. I. Shrira and V. V. Geogjaev, What makes the Peregrine soliton so special as a prototype of freak waves?, \href{https://link.springer.com/article/10.1007/s10665-009-9347-2}{J. Eng. Math. 67, 11 (2010)}.
\bibitem{pattern2009} N. Akhmediev, A. Ankiewicz, and J. M. Soto-Crespo, Rogue waves and rational solutions of the nonlinear Schr\"odinger equation, \href{https://journals.aps.org/pre/abstract/10.1103/PhysRevE.80.026601}{Phys. Rev. E 80, 026601 (2009).}
\bibitem{theoryreview2017} N. Akhmediev, A. Ankiewicz, and J. M. Soto-Crespo, Fundamental rogue waves and their superpositions in nonlinear integrable systems, In: S. Wabnitz, (Ed.), \emph{Nonlinear Guided Wave Optics: A testbed for extreme waves}, (IOP Publishing, Bristol, 2017).
\bibitem{review2013} M. Onorato, S. Residori, U. Bortolozzo, A. Montina, and F. T. Arecchi, Rogue waves and their generating mechanisms in different physical contexts, \href{https://www.sciencedirect.com/science/article/pii/S0370157313000963}{Phys. Rep. \textbf{528}, 47 (2013)}.
\bibitem{review2014} J. M. Dudley, F. Dias, M. Erkintalo, and G. Genty, Instabilities, breathers and rogue waves in optics, \href{https://www.nature.com/articles/nphoton.2014.220}{Nat. Photonics \textbf{8}, 755 (2014)}.
\bibitem{review2019} J. M. Dudley, G. Genty, A. Mussot, A. Chabchoub, and F. Dias, Rogue waves and analogies in optics and oceanography, \href{https://doi.org/10.1038/s42254-019-0100-0}{Nat. Rev. Phys. 1 675 (2019).}
\bibitem{Peregrine} D. H. Peregrine, Water waves, nonlinear Schr\"odinger equations and their solutions. {\it J. Aust. Math. Soc. Ser. B} \textbf{25}, 16-43 (1983).
\bibitem{RWO-1} B. Kibler, J. Fatome, C. Finot, G. Millot, F. Dias, G. Genty, N. Akhmediev, and J. M. Dudley, The Peregrine soliton in nonlinear fibre optics, \href{https://doi.org/10.1038/nphys1740}{Nature Phys. 6, 790 (2010)}.
\bibitem{RWO-2} A. Chabchoub, S. Neumann, N. P. Hoffmann, and N. Akhmediev, Spectral properties of the Peregrine soliton observed in a water wave tank, J. Geophys. Res. 117, C00J03 (2012).
\bibitem{Spectra2011-1} N. Akhmediev, A. Ankiewicz, J. M. Soto-Crespo, and J.M. Dudley, Rogue wave early arning through spectral measurements?,  \href{https://doi.org/10.1016/j.physleta.2010.12.027}{Phys. Lett. A, \textbf{375}, 541 (2011)}.
\bibitem{Spectra2011-2} N. Akhmediev, J. M. Soto-Crespo, A. Ankiewicz, and N. Devine, Early detection of rogue waves in a chaotic wave field, Phys. Lett. A 375, 2999 (2011).
\bibitem{PLA2012} U. Bandelow and  N. Akhmediev, Persistence of rogue waves in extended nonlinear Schr\"odinger equations: Integrable Sasa-Satsuma case, Phys. Lett. A, {\bf 376}, 1558 (2012).
\bibitem{ds-oyI} Y. Ohta and J. Yang, Rogue waves  in the Davey-Stewartson I equation, Phys. Rev. E {\bf 86}, 036604, (2012).
\bibitem{FB3} F. Baronio, M. Conforti, A. Degasperis, and S. Lombardo, Rogue waves emerging from the resonant interaction of three waves, \href{https://journals.aps.org/prl/abstract/10.1103/PhysRevLett.111.114101}{Phys. Rev. Lett. 111, 114101 (2013)}.

\bibitem{JS2013} C. Z. Li, J. S. He, and K. Porsezian, Rogue waves of the Hirota and the Maxwell-Bloch equations, Phys. Rev. E 87, 012913 (2013).
\bibitem{VRW-chap-2016} A. Degasperis and S. Lombardo, \emph{Integrability in action: Solitons, instability and rogue waves}, In:  M. Onorato, S. Residori, and F. Baronio, ed., \textit{Rogue and Shock Waves in Nonlinear Dispersive Media} \href{https://doi.org /10.1007/978-3-319-39214-1_2 }{(Springer, 2016)}.



\bibitem{Feng2016} L. Ling, B.-F. Feng, and Z. Zhu, Multi-soliton, multi-breather and higher order rogue wave solutions to the complex short pulse equation, Physica D 327, 13 (2016).

\bibitem{Rewiew-2017} S. Chen, F. Baronio, J. M. Soto-Crespo, Ph. Grelu, and, D. Mihalache, Versatile rogue waves in scalar, vector, and multidimensional nonlinear systems, \href{https://iopscience.iop.org/article/10.1088/1751-8121/aa8f00/meta}{J. Phys. A: Math. Theor. {\bf 50}, 463001 (2017)}.
\bibitem{Vobservation1} B. Frisquet, B. Kibler, P. Morin, F. Baronio, M. Conforti, G. Millot, and S. Wabnitz, Optical dark rogue waves, \href{https://doi.org/10.1038/srep20785}{Sci. Rep., {\bf 6}, 20785 (2016)}.
\bibitem{Vobservation2} F. Baronio, B. Frisquet, S. Chen, G. Millot, S. Wabnitz, and B. Kibler, Observation of a group of dark rogue waves in a telecommunication optical fiber, \href{https://doi.org/10.1103/PhysRevA.97.013852}{Phys. Rev. A {\bf 97}, 013852 (2018)}.
\bibitem{Zhao-2014} L.-C. Zhao, G.-G. Xin, Z.-Y. Yang,  Rogue-wave pattern transition induced by relative frequency, \href{https://doi.org/10.1364/OE.17.021497}{Phys. Rev. E 90, 022918  (2014)}.
\bibitem{Akhmediev-2015} N. Akhmediev, J. M. Soto-Crespo, N. Devine, and N. P. Hoffmann, Rogue wave spectra of the Sasa-Satsuma equation, Physica D 294, 37 (2015).
\bibitem{MM} S. V. Manakov, On the theory of two-dimensional stationary self-focusing of electromagnetic waves, \href{http://jetp.ac.ru/cgi-bin/dn/e_038_02_0248.pdf}{Sov. Phys. JETP, {\bf 38}, 248 (1974)}.
\bibitem{N-RW} C. Liu, S.-C. Chen, X. Yao, and  N. Akhmediev, Non-degenerate multi-rogue waves and easy ways of their excitation, \href{https://doi.org/10.1016/j.physd.2022.133192}{Physica D: Nonlinear Phenomena 433, 133192 (2022)}.
\bibitem{BEC} P. G. Kevrekidis, D. Frantzeskakis, and R. Carretero- Gonzalez, Emergent nonlinear phenomena in Bose-Einstein condensates: Theory and experiment \href{https://doi.org/10.1007/978-3-540-73591-5}{(Springer, Berlin Heidelberg, 2009)}.
\bibitem{OF} G. Agrawal, \emph{Nonlinear Fiber Optics}, \href{https://www.sciencedirect.com/book/9780123970237/nonlinear-fiber-optics}{5th ed. (Academic Press, San Diego, 2012)}.
\bibitem{F} M. Onorato, A. R. Osborne, and M. Serio, Modulational instability in crossing sea states: A possible mechanism for the formation of freak waves, \href{https://doi.org/10.1103/PhysRevLett.96.014503}{Phys. Rev. Lett., {\bf 96}, 014503 (2006)}.
\bibitem{DT1991} V. B. Matveev and M. A. Salle, \textit{Darboux Transformations and Solitons}, Series in Nonlinear Dynamics. (Springer Verlag, Berlin, 1991).
\bibitem{VMRW3} L. Ling, L.-C. Zhao, Z. Yang, and B. Guo, Generation mechanisms of fundamental rogue wave spatial-temporal
structure, \href{https://journals.aps.org/pre/abstract/10.1103/PhysRevE.96.022211}{Phys. Rev. E 96, 022211 (2017)}.
\bibitem{Zhao-2012} L.-C. Zhao and J. Liu, Localised nonlinear waves in a two-mode nonlinear fiber, \href{https://doi.org/10.1364/josab.29.003119}{J. Opt. Soc. Am. B {\bf 29}, 3119 (2012)}.
\bibitem{Zhao-2013} L.-C. Zhao and J. Liu, Rogue-wave solutions of a three-component coupled nonlinear Schr\"odinger equation, \href{https://doi.org/10.1103/physreve.87.013201}{Phys. Rev. E {\bf 87}, 013201 (2013)}.
\bibitem{Baronio-2014} F. Baronio, M. Conforti, A. Degasperis, S. Lombardo, M. Onorato, and S. Wabnitz, Vector rogue waves and baseband modulation instability in the defocusing regime \href{https://doi.org/10.1103/PhysRevLett.113.034101}{Phys. Rev. Lett., {\bf 113}, 034101 (2014)}.
\bibitem{Qin-2023} Y.-H. Qin, L. Ling, and L.-C. Zhao, Optical rogue-wave patterns in coupled defocusing systems, \href{https://doi.org/10.1103/PhysRevA.108.023519}{Phys. Rev. A 108, 023519 (2023)}.
\bibitem{VAB2021} S.-C. Chen, C. Liu, X. Yao, L.-C. Zhao, and N. Akhmediev, Extreme spectral asymmetry of Akhmediev breathers and Fermi-Pasta-Ulam recurrence in a Manakov system, \href{https://doi.org/10.1103/PhysRevE.104.024215}{Phys. Rev. E {\bf 104}, 024215 (2021)}.
\bibitem{VAB2022} C. Liu, S.-C. Chen, X. Yao, and N. Akhmediev, Modulation instability and non-degenerate Akhmediev breathers of Manakov equations,
\href{https://iopscience.iop.org/article/10.1088/0256-307X/39/9/094201}{Chin. Phys. Lett., {\bf 39}, 094201 (2022)}.
\bibitem{VAB-Df2022} S.-C. Chen and C. Liu, Hidden Akhmediev breathers and vector modulation instability in the defocusing regime, \href{https://doi.org/10.1016/j.physd.2022.133364}{Physica D {\bf 438}, 133364 (2022)}.
\end{thebibliography}
\end{document}